\documentclass[conference]{IEEEtran}
\IEEEoverridecommandlockouts
\usepackage[T1]{fontenc}
\usepackage{gensymb}
\usepackage[utf8]{inputenc}
\usepackage{pdfpages}
\usepackage{booktabs}
\usepackage{multirow}
\usepackage{lscape}
\usepackage{booktabs,caption}
\usepackage[flushleft]{threeparttable}
\usepackage{array}
\usepackage{bm}
\usepackage{booktabs}
\usepackage{cite}
\usepackage{amsmath,amssymb,amsfonts}
\usepackage{algorithmic}
\usepackage{graphicx}
\usepackage{textcomp}
\usepackage{xcolor}
\ifCLASSOPTIONcompsoc
  \usepackage[caption=false,font=normalsize,labelfont=sf,textfont=sf]{subfig}
\else
  \usepackage[caption=false,font=footnotesize]{subfig}
\fi
\def\BibTeX{{\rm B\kern-.05em{\sc i\kern-.025em b}\kern-.08em
    T\kern-.1667em\lower.7ex\hbox{E}\kern-.125emX}}
\begin{document}

\title{Air-to-Ground Directional Channel Sounder With 64-antenna Dual-polarized Cylindrical Array

\thanks{This material is based upon work supported by the National Science Foundation under Grant No. 1923601. J. Gomez-Ponce's research work is partially supported by the Foreign Fulbright Ecuador SENESCYT Program}
}

\author{\IEEEauthorblockN{Jorge Gomez-Ponce\IEEEauthorrefmark{1}\IEEEauthorrefmark{3}, Thomas Choi\IEEEauthorrefmark{1}, Naveed A. Abbasi\IEEEauthorrefmark{1}, Aldo Adame\IEEEauthorrefmark{2}, Alexander Alvarado\IEEEauthorrefmark{2},\\ Colton Bullard\IEEEauthorrefmark{1}, Ruiyi Shen\IEEEauthorrefmark{1}, Fred Daneshgaran\IEEEauthorrefmark{2}, Harpreet S. Dhillon\IEEEauthorrefmark{4} and Andreas F. Molisch\IEEEauthorrefmark{1}}
\IEEEauthorblockA{\IEEEauthorrefmark{1}
University of Southern California, Los Angeles, CA, USA \\
Email:\{gomezpon, choit, nabbasi, ctbullar, ruiyishe, molisch\}@usc.edu}
\IEEEauthorblockA{\IEEEauthorrefmark{2}
California State University, Los Angeles, CA, USA \\
Email:\{aadame14, aalva203, fdanesh\}@calstatela.edu}
\IEEEauthorblockA{\IEEEauthorrefmark{3}ESPOL Polytechnic University, Escuela Superior Politécnica del Litoral, ESPOL,\\ Facultad de Ingenier\'ia en Electricidad y Computaci\'on, Km 30.5 vía Perimetral, P. O. Box 09-01-5863, Guayaquil, Ecuador\\}
\IEEEauthorblockA{\IEEEauthorrefmark{4} Wireless@VT, Bradley Department of Electrical and Computer Engineering, Virginia Tech, USA\\
Email:hdhillon@vt.edu}
}

\maketitle

\begin{abstract}
Unmanned Aerial Vehicles (UAVs), popularly called drones, are an important part of future wireless communications, either as user equipment that needs communication with a ground station, or as base station in a 3D network. For both the analysis of the "useful" links, and for investigation of possible interference to other ground-based nodes, an understanding of the air-to-ground channel is required. Since ground-based nodes often are equipped with antenna arrays, the channel investigations need to account for it. This study presents a massive MIMO-based air-to-ground channel sounder we have recently developed in our lab, which can perform measurements for the aforementioned requirements. After outlining the principle and functionality of the sounder, we present sample measurements that demonstrate the capabilities, and give first insights into air-to-ground massive MIMO channels in an urban environment. Our results provide a platform for future investigations and possible enhancements of massive MIMO systems.
\end{abstract}


\section{Introduction}
Massive  multiple-input multiple-output (MIMO) systems aim to use a large number of antennas at the base stations (BSs), usually on the order of tens to hundreds, to increase the spectral and energy efficiency of wireless networks \cite{marzetta2016fundamentals,bjornson2017massive}. These characteristics make massive MIMO systems attractive for 5G and beyond wireless communication systems. However, characterization of a wireless system by means of channel sounding is essential before an eventual deployment is considered. Therefore, extensive channel measurements for massive MIMO channels in various scenarios of interest are very important. Considering this, we have investigated massive MIMO channels in a number of our previous studies, such as \cite{choi2019channel,rottenberg2020experimental,choi2020co}, using a real-time channel sounder with a 64-antenna dual-polarized cylindrical array that operates at 3.5 GHz.  

Unmanned aerial vehicles (UAVs), also known as drones and unmanned aerial systems (UAS), have seen phenomenal growth worldwide for a wide and expanding variety of applications \cite{matolak2015unmanned}. Traditionally a tool of military and hobbyists that was quite expensive, the advancement of both fixed-wing and, more significantly, rotorcraft airframes along with comprehensive flight control technologies has brought drone use to many including consumers, public safety organizations, scientists and engineers, and telecommunications companies \cite{bor20195g}. From a communication perspective,
drone-to-ground links are both interesting for communication of ground stations with the drones, and for the case where the drone serves as infrastructure in a 3D network; the latter case is attractive due to drones' ability to enhance the capacity, coverage and reliability for existing networks on-demand with low cost and infrastructure overheads \cite{bekmezci2013flying}. In the latter case, communication from the drone to a ground BS might be of interest as a backhaul link, or to assess the interference that a drone might have on an existing BS intended for traditional land mobile services. Finally, drones might also communicate with ground-based user equipments (UEs) that have multiple antenna elements, e.g., mounted on cars or trucks.  

A number of studies on UAV and drone networks have looked at the air-to-ground (A2G) communication channels, see the survey \cite{khawaja2019survey} and references therein. Several recent studies have motivated the use of massive MIMO communications with drone swarms due to their potential for enhancements of current and future networks \cite{chandhar2017massive,geraci2018understanding}. As we discussed earlier, extensive channel sounding is required before the eventual deployment of a wireless system is possible. Therefore, in the current study we discuss the development of an A2G directional massive MIMO channel sounder. Exploiting components of our existing real-time massive MIMO sounder for ground-based communications end, we add a newly developed remote drone communication end and refine synchronization and postprocessing for the new application. With the new sounder, we execute basic performance tests and perform and evaluate a sample measurement campaign in an urban scenario. Our results provide a platform for further channel measurements of drone-based extensions of massive MIMO systems and their eventual large-scale development.

The rest of this paper is organized as follows. In section II, we discuss the channel sounder hardware. The measurement scenario is described in section III. Results of the sounder are presented in section IV and the manuscript is finally concluded in section V. 

\section{Channel Sounder Hardware}
\begin{figure}[!t]
    \centering
    \subfloat[Aerial transmitter on drone.]{\includegraphics[width=0.95\linewidth]{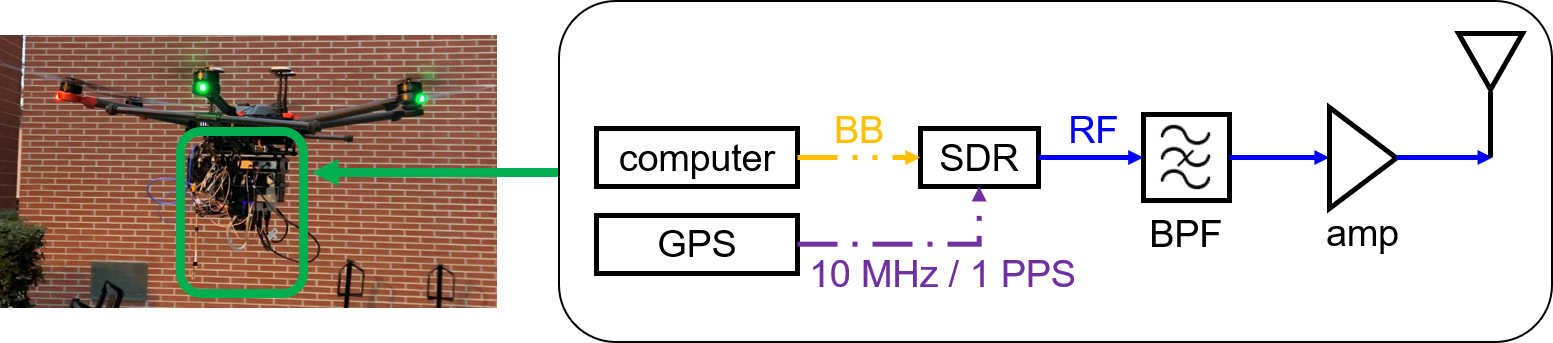}%
    \label{sndr_drone}}
    \newline
    \subfloat[Ground receiver with 64-antenna dual-polarized cylindrical array.]{\includegraphics[width=0.95\linewidth]{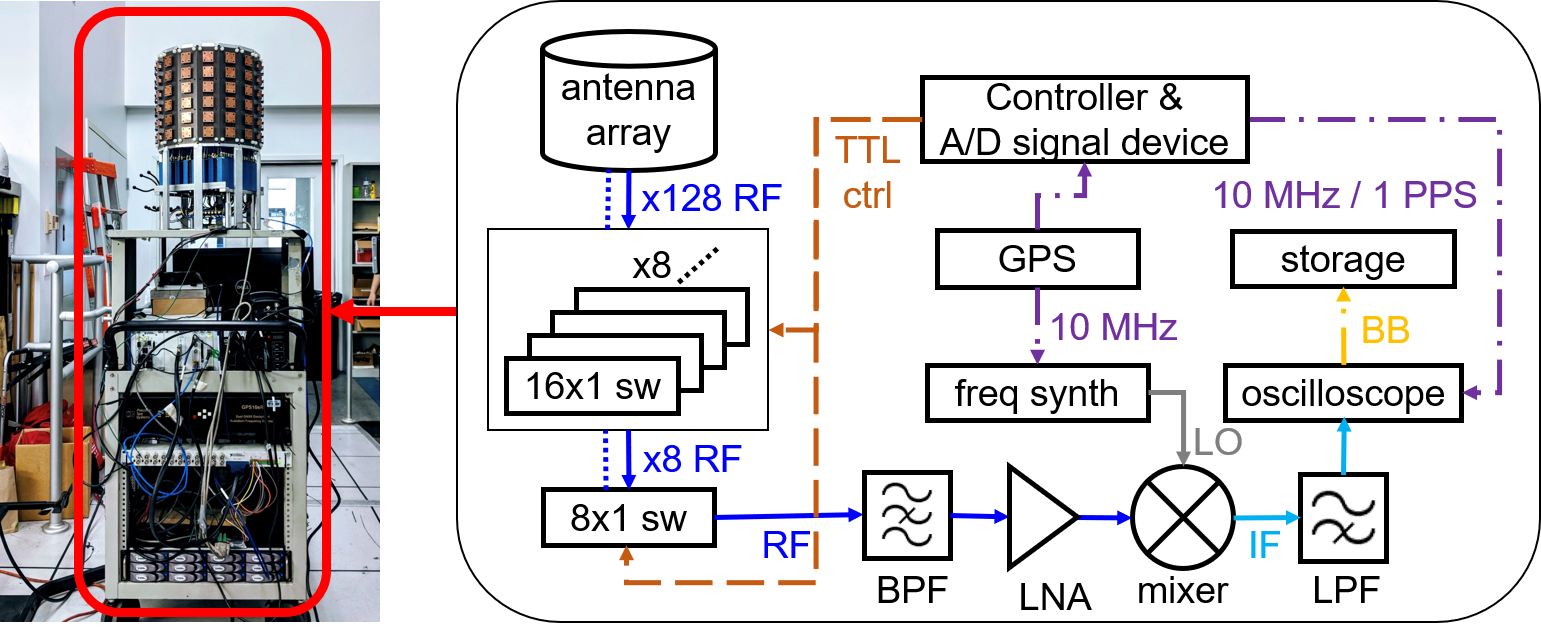}%
    \label{sndr_mMIMO}}
    \caption{A2G directional channel sounder hardware.}
    \label{sndr}
\end{figure}

In this section, we introduce our A2G channel sounder to measure single-input, multi-output (SIMO) massive MIMO scenarios.\footnote{While the use of massive "MIMO" seems self-contradictory when one link end has only a single antenna, this nomenclature is rooted in the anticipated use of multiple single-antenna terminals, and is well established in the literature.} The channel sounder is composed of an aerial transmitter (TX) on a drone and a ground receiver (RX) with 64-antenna dual-polarized massive MIMO cylindrical array, as shown in Fig. \ref{sndr}. Due to the channel reciprocity, measurements taken in this configuration are equally valid for a ground (TX) - to -drone (RX) link. 
A table containing the hardware list of the major sounder components is shown in Table \ref{tab:Hardware_list}.

\vspace{0.1in}
\begin{table}[t]
\centering
\caption{Major hardware components.}
\label{tab:Hardware_list}
\begin{tabular}{|l|l|l|}
\hline
\textbf{Item} & \textbf{Manufacturer} & \textbf{Model No.} \\ \hline
Drone         & DJI                  & DJI Matrice 600 Pro \\ \hline
USRP          & National Instruments & NI2901              \\ \hline
GPS TX        & Jackson Labs         & Firefly 1A          \\ \hline
Controller TX & Intel                & NUC10i7FNK          \\ \hline
BPF           & K\&L Microwave       & 3C50-3500/T50-O/O   \\ \hline
Switch        & Pulsar Microwave     & SW16AD-21/SW8AD-A85 \\ \hline
Mixer         & Mini Circuits        & ZEM-4300MH+         \\ \hline
FS            & Phase Matrix         & FSW-0020            \\ \hline
PA            & Wenteq               & ABP1500-03-3730     \\ \hline
LNA           & Wenteq               & ABL0600-33-4009     \\ \hline
LPF           & Pasternack           & PE8719              \\ \hline
Controller RX & National Instruments & PXIe-1082           \\ \hline
GPS RX        & PTSYST               & GPS10eR             \\ \hline
\end{tabular}
\end{table}

\subsection{Aerial transmitter: Drone}
A ``DJI Matrice 600 Pro'' drone is used as a carrier of the TX payloads. This drone has a dimension of 1.67 m $\times$ 1.52 m $\times$ 0.73 m, and can carry a payload of up to  5.5 kg. In our current design, the sounder payload includes a minicomputer that creates our baseband signal (Intel NUC 10), a software-defined radio (SDR) transmitting 46 MHz bandwidth signal sampled at 50 MS/s centered around 3.5 GHz carrier frequency (NI USRP-2901), a 50 MHz band-pass filter (K\&L Microwave 3C50-3500/T50-O/O), a 30 dBm amplifier (Wenteq Microwave ABP1500-03-3730), an omni-directional antenna (Decawave WB002), a global positioning system (GPS) which synchronizes the TX and the RX using 1 pulse-per-second (PPS) signal and provides frequency reference to the SDR using 10 MHz signal (Jackson Labs Firefly-1A), and a portable battery (Powkey 200Watt) for powering this payload. All the components that form the payload are safely mounted on a custom-built rack that is designed to prevent payloads from accidentally falling off. The selection of the sounder components in terms of bandwidth, output power, number of antennas, and clock accuracy was done in such a way that the payload weight could be minimal and thus the flight time can be extended. With the current settings, our setup is able to provide about 15 minutes of continuous real-time measurements per battery charge.
Additionally, a common problem when doing measurement on a route is to track the location of the transmitter or receiver in every location. Our design allows the setting of predefined routes for the drone thereby increasing the precision of measurements in specific locations and routes in contrast to any measurements conducted by manual movement of transmitter or receiver.
\vspace{0.1in}

\subsection{Ground receiver: 64-antenna dual-polarized cylindrical array}
On the ground side, a \textit{fast-switching array system} which connects one antenna port at a time to the receiver downconversion chain is used. The custom-built cylindrical antenna array is created by 16 linear antenna arrays (which we refer to as columns) forming a circle. Each column has six parasitic patch antenna elements, and only the four center elements are used during the measurement, while top and bottom elements serve as dummies that improve uniformity of mutual-coupling effects. Each antenna element contains two ports that receive vertically and horizontally polarized radiation, respectively. Therefore, the array contains 16 $\times$ 4 $\times$ 2 = 128 antenna ports, with 360$^{\circ}$ field of view in azimuth and 120$^{\circ}$ field of view in elevation. The beam pattern of the array is shown in \cite{choi2019channel}. The sounder switches every 50$\:\mu$s, resulting in 6.4 ms per SIMO measurement.

Aside from the antenna array, the RX contains 8 sets of 16$\times$1 switches connected to all antenna ports of the array (Pulsar Microwave SW16AD-21), an 8$\times$1 switch connected to 8 outputs of 16$\times$1 switches (Pulsar Microwave SW8AD-A85), a 40 dB gain low noise amplifier (Wenteq ABL0600-33-4009), a mixer for downconversion (Mini-Circuits ZEM-4300MH+), a frequency synthesizer for local oscillator signal (Phase Matrix FSW-0020), a 100 MHz bandwidth low-pass filter at the 50 MHz to 96 MHz intermediate frequency (Pasternack PE8719), an oscilloscope converting analog intermediate frequency signals to digital baseband frequency signals at 250 MS/s (NI PXIe 5160), a storage for received data (NI HDD-8265), a GPS for TX/RX synchronization 1 PPS signal and frequency reference 10 MHz signal (Precision Test Systems GPS10eR), a controller and an analog/digital signal input/output device to control the switches with transistor-transistor logic (TTL) signal and distribute the GPS outputs to the oscilloscope (NI PXIe 8135 and NI BNC-2090A), and a battery bank to power RX hardware in outdoor environments (Be Prepared Solar 20L-600). All these components are mounted on a mobile cart, which can be placed at a selected location during a drone flight, or be moved at a walking speed while the drone hovers.
\vspace{0.1in}
\begin{table}[]
\centering
\caption{Key sounder parameters.}
\label{tab:parameters}
\begin{tabular}{@{}|l|l|@{}}
\hline
\textbf{Parameter}                & \textbf{Value} \\ \hline
Carrier Frequency                 & 3.5 GHz        \\ \hline
Bandwidth                         & 46 MHz         \\ \hline
Transmit power                    & 27 dBm          \\ \hline
Sampling rate TX                  & 50 MS/s        \\ \hline
Sampling rate RX                  & 250 MS/s       \\ \hline
SISO signal duration              & 50 $\mu s$     \\ \hline
SIMO duration                     & 6.4 ms         \\ \hline
Number of frequency points (RX)   & 1841           \\ \hline
Number of TX antennas             & 1              \\ \hline
Number of RX antennas             & 128            \\ \hline
Number of SIMO captures per burst & 3              \\ \hline
Rate of burst                     & 20 Hz          \\ \hline
\end{tabular}
\end{table}

\subsection{Sounding signal design, processing and stationary behavior of the system}
A summary of the sounder configuration parameters is given in Table \ref{tab:parameters}. The procedure used to generate the sounding signal is explained in \cite{UAV_Survey,V2V}. As result of it an OFDM signal of duration $T_{SISO}=50\mu s$ \footnote{Given limitations of the hardware (e.g. USRP), Doppler analysis is not perform at the current stage. Drone speed is set to be $\leq 3m/s$. $T_{SISO}$ was selected as the minimum value in which the USRP performs without underrun errors} is sent to the USRP at $50MS/s$ to be streamed to the RX.\\
To characterize the frequency response of the system it is necessary to perform back-to-back (B2B) calibrations on it. To perform this, we replace the connections to the antennas with an RF cable and measure the response of the system.  A well characterized attenuator is used in this process to avoid any damage to the receiver.
$Y_{B2B}$ is the B2B calibration that will be used for post-processing, however it is necessary to eliminate the response of the attenuator used in the B2B calibration. To obtain the final "antenna+channel" response for the post-processing we follow the procedure in \cite{V2V},
\begin{equation} \label{eq:2}
    H(f)=\frac{Y_{meas}(f)}{Y_{ref}(f)}\times G_{att}(f),
\end{equation}
where $H(f)$ is the "antenna+channel" response, $Y_{meas}(f)$ is the measured transfer function and $G_{att}(f)$ is the frequency response of the attenuator used during the B2B measurement.

\begin{figure}[!t]
    \begin{center}
    \subfloat[Magnitude.]{\includegraphics[width=0.8\linewidth]{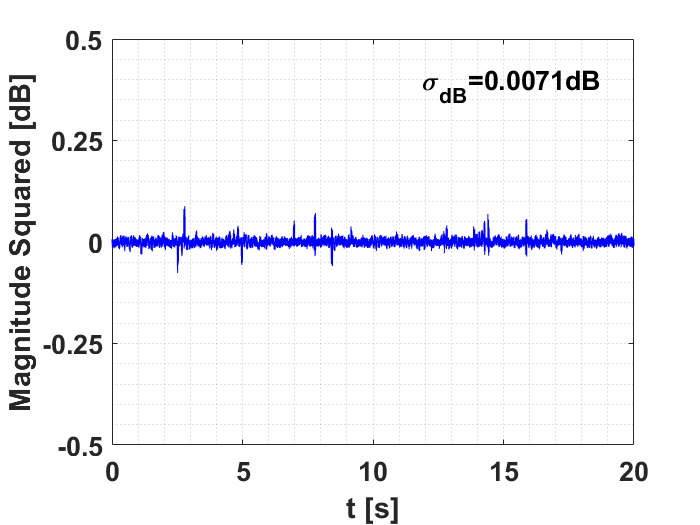}%
    \label{mag_B2B}}    
    \end{center}
    \begin{center}
    \subfloat[Phase.]{\includegraphics[width=0.8\linewidth]{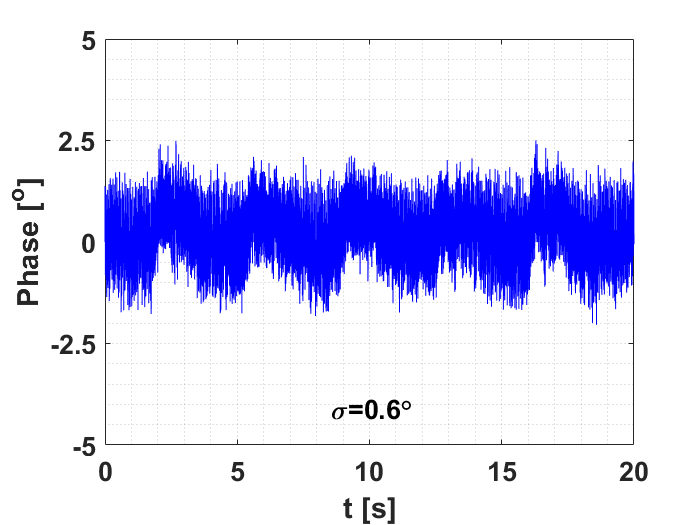}%
    \label{pha_B2B}}   
    \end{center}
    \caption{Stability of the B2B results using a shared clock reference.}
    \label{B2B}
\end{figure}

An important test for any sounder is to verify the consistency of its results in a long measurement. For this, we perform a 20 second long B2B measurement with a shared clock signal to verify the stability of the sounder. Significant phase variation in the same SIMO snapshot would compromise the accuracy of any high resolution parameter estimation (HRPE) algorithm that may be used to eventually process the data in more detail.
In Fig. \ref{B2B}, the relative amplitude and phase responses with respect to the first SISO snapshot are shown. 
The standard deviation observed for the amplitude and phase over the 20 second period was just $0.0071$ dB and $0.6^\circ$, respectively. Note that this experiment shows the stability of the sounder itself based on a single clock. When using different clocks on both ends of our system, the accuracy can change, as will be discussed below. 

\section{Scenario description}
As part of the current study we conducted an extensive channel measurement campaign for various A2G channel scenarios, such as: LOS, NLOS, around-the-corner, multiple heights etc. So far, the total number of SIMO snapshot we measured for A2G environments is over 60000, resulting in approximately 2 TB of data at the moment.
In this section, we describe the measurement scenarios and present sample results of the post-processed data obtained from the sounder.

\vspace{0.1in}
\begin{figure}[!t]
    \centering
    \includegraphics[width=0.8\linewidth]{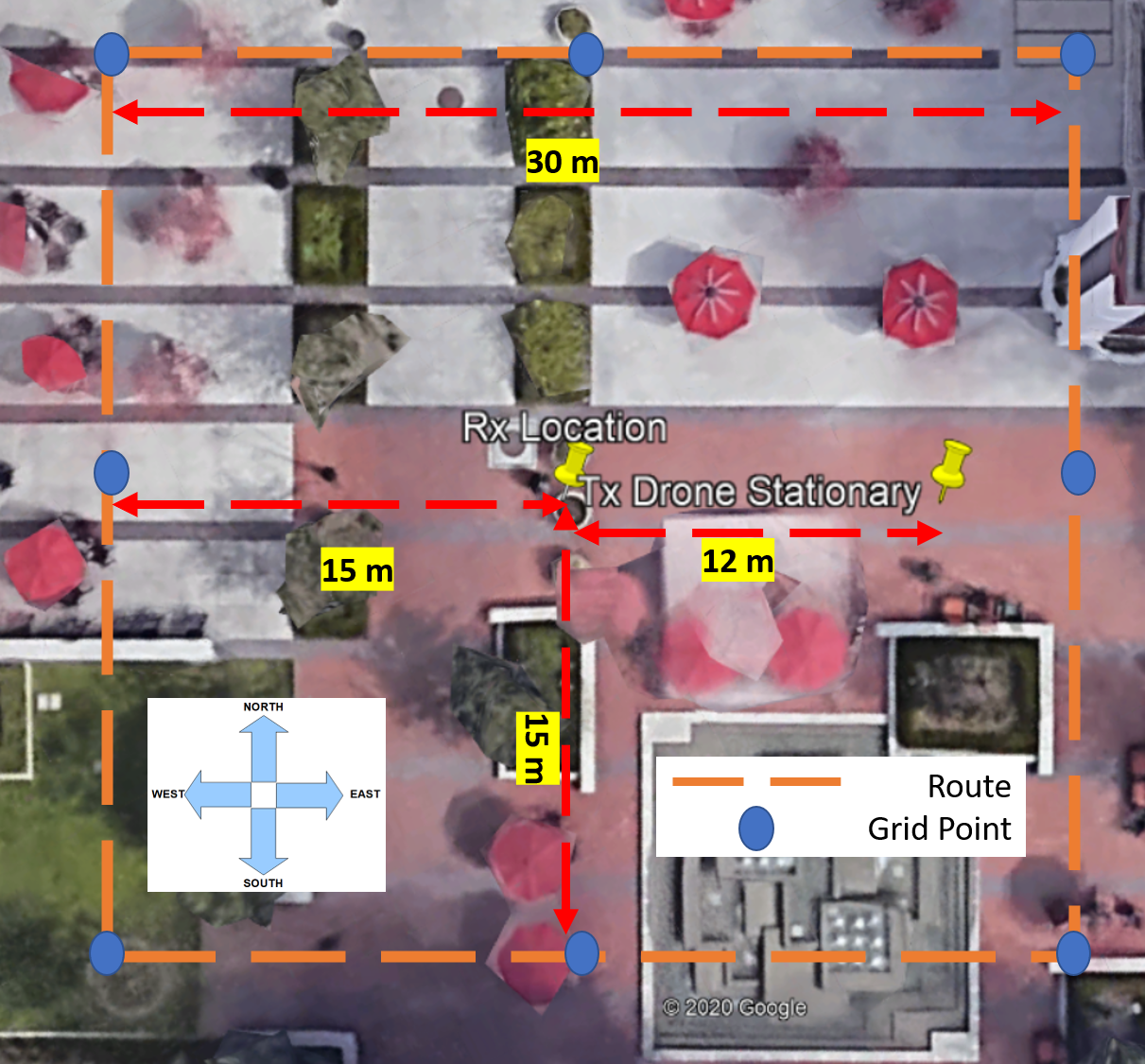}
    \caption{Measurement scenario.}
    \label{scenario}
\end{figure}

To validate the sounder, we perform an experiment comparing our sounder in a completely stationary case (mounted on a pole) versus the drone hovering over the same location. For this case, the measurement was performed in the quadrangular courtyard in front of Olin Hall of Engineering on the University of Southern California (USC)  University Park Campus, Los Angeles. The area is surrounded by tables, umbrellas, buildings, trees and a water fountain (see Fig. \ref{scenario}).
For this measurement, distance between TX and RX was approximately 12 meters and the antenna of the transmitter was set to be 1.8 meters from the ground. The orientation of the RX is such that column 1 of antennas face south of the drone, columns 5 faces east, column 9 faces north and column 13 looks west. Given the location of the drone, east from the RX, column 5 is the one facing to the drone.  
\subsection {Sample LOS route}
In the second experiment, we move a drone on a route at a specific speed. The location is the same as for the first experiment. The route is a 30 meter square with the RX located approximately at the center of the square. The height of the drone is 50 meters from the ground. For this case the speed of the drone was set to be 2 m/s (see Fig. \ref{scenario}). The orientation of the drone is similar to the first experiment, the drone starts at the north west corner of the route, facing columns 9 and 10, going from west to east, then north to south, going east to west and finally completing the route.  
\section{Results and Discussion}
\subsection{Processing Procedure}
Given the scenario and experiments previously described, our focus will be the comparison between the different scenarios. In our measurements, we send copies of the RF signal that are then received by each of the RF ports of the antenna array. As a result of this process, a collection of RF signals that correspond to the captures of all RF ports is received. In order to eliminate the effects of the system, a B2B calibration is performed to ensure the signal only contains the "antenna+channel" behavior, as explained in (\ref{eq:2}). We then compute the channel impulse response (CIR) for each port element k, $h_k' (\tau)$. The impulse response is obtained from the transfer function via inverse Fourier transform. Subsequently, noise thresholding and delay gating is performed, similar to the description in \cite{gomez-ponce2020}. The larger of two noise thresholds, ($P_\lambda$), is applied where the first is set to 6dB above the noise floor and the other as 20dB below the peak power; the time for delay gating is $\tau_\lambda=2\mu s$. 
\\
Using the modified impulse response $h_k (\tau)$, we proceed to compute the following parameters:
\begin{enumerate}
    \item The total RX power per drone location (i), as the non-coherent addition of the contribution of each antenna port \begin{equation}
        P_{RX_i}=\sum_{k=1}^K |h_{k,i}(\tau)|^2.
    \end{equation}
    \item A second parameter for evaluation is the root mean squared (RMS) delay spread for the strongest antenna element, computed according to \cite{molisch2012wireless}, Chapter 6.
    \item The next parameter to be used is the correlation matrix, given the "channel+antenna" response we compute it as follows:
    \begin{equation}
        \mathbf{R_i}=E_f\{\mathbf{H}(f)_{stack,i}\mathbf{H}(f)_{stack,i}^\dagger\},
    \end{equation}
    where $\mathbf{H}(f)_{stack,i}=[H(f)_{1,i}\cdots H(f)_{N_{RX},i}]^T$ is the "stacked" channel frequency response in location "i", $H^T$ and $H^\dagger$ are the transpose and conjugate transpose of $H$ respectively, $H(f)_{l,i}$ is the channel frequency response of location $i$ at frequency point $f$ and RX antenna $l$, and $E_f$ represents the expectation over frequency. Due to the larger coherence bandwidth, the number of independent samples in the frequency domain is low, so that $R$ is not a valid correlation matrix in the traditional sense, but still does provide information about diversity efficiency.
    From $\mathbf{R_i}$, we compute the distribution of the sorted eigenvalues and, in particular, the ratio between the first and second, and first and fourth eigenvalues $\gamma_{12}(R_i)=\frac{E_1}{E_2},\gamma_{14}(R_i)=\frac{E_1}{E_4}$ where $E_n$ is the $n^{th}$ eigenvalue of the matrix $R_i$.
\end{enumerate}

The procedure explained before will be used for all scenarios.
\subsection{Results Discussion}
\begin{figure}[!t]
    \begin{center}
    \subfloat[Static.]{\includegraphics[width=0.8\linewidth]{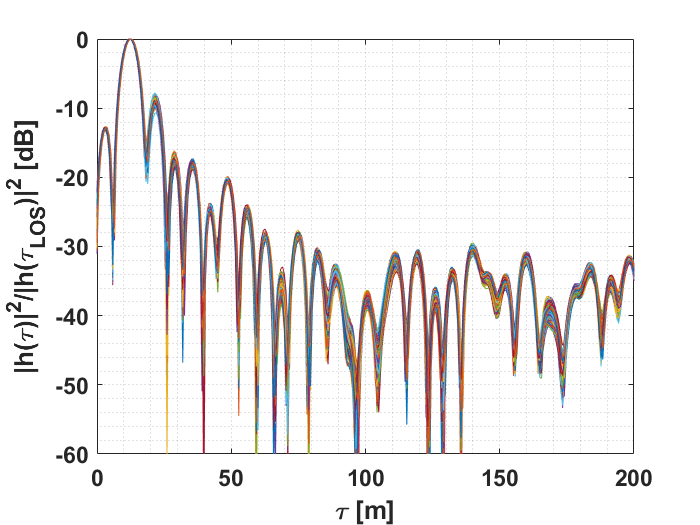}%
    \label{static_PDPs}}    
    \end{center}
    \begin{center}
    \subfloat[Hover.]{\includegraphics[width=0.8\linewidth]{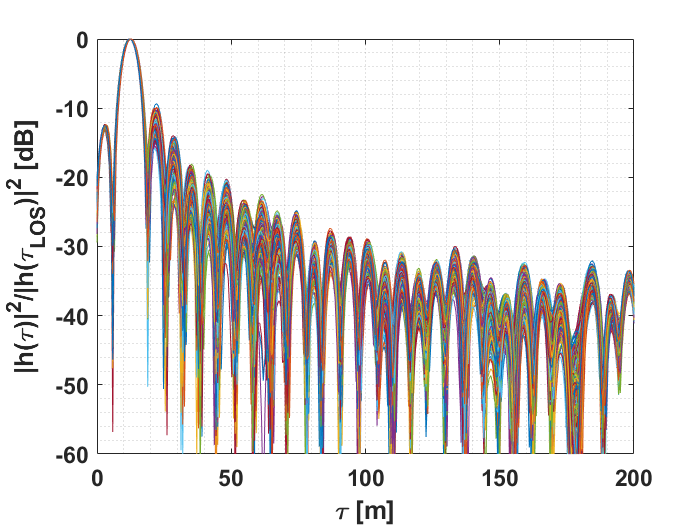}%
    \label{hover_PDPs}}    
    \end{center}
    \caption{Overlapping normalized PDP strongest antenna port facing towards drone.}
    \label{PDPs}
\end{figure}
\begin{figure}[!t]
    \begin{center}
    \subfloat[Static.]{\includegraphics[width=0.8\linewidth]{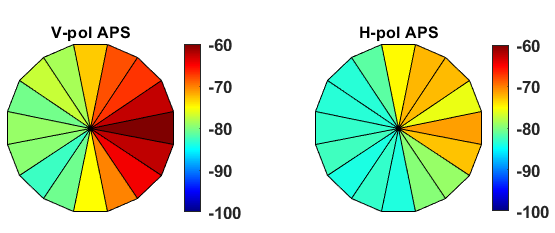}%
    \label{static_angularplot}}        
    \end{center}
    \begin{center}
    \subfloat[Hover.]{\includegraphics[width=0.8\linewidth]{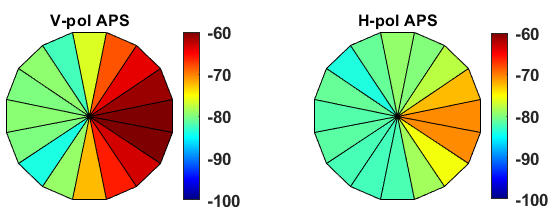}%
    \label{hover_angularplot}}    
    \end{center}
    \caption{Receiver Angular power distribution per column of a single snapshot}
    \label{angularplot}
\end{figure}

From physical considerations, we anticipate that variations of the receive power will be higher in the hovering scenario because of the vibrations of the drone and the inability of hovering to keep the drone in {\em exactly} the same location. In Fig. \ref{PDPs}, we show normalized PDPs of both these cases for different captures of the sounder. It can be observed that hovering clearly has more variations than the static case.\\
Looking at the variations of the LOS  bin power, it confirms the conjecture that the hovering creates additional variations compared to the static case: the standard deviation for the LOS bin power is 0.08dB for the static case and 0.48dB for hovering. This behavior is consistent with the insights in \cite{UAV_wob}, where a Rice-fading channel in the presence of wobbling of a drone was analyzed. 
This work shows that even small movements of the drone can lead to time variations of the received power (behavior that is observed in our experiments) and so decreases the channel coherence time.\\   
Comparing the 2 polarizations, we expect that the highest power will be on the "Vertical" polarization compared to the "Horizontal" one, given the fact that the Tx antenna is vertically polarized. Figure \ref{angularplot} shows the angular distribution of a single snapshot for both the hovering and static case.

In both cases we see there is concentration of power on the right which is the direction in which the drone is located. Furthermore, we see that the power in the V-polarization is approximately 12dB higher than in the H-polarization. It is important to mention that the antenna was placed in the same orientation as it was for the hovering experiment (upside down) using a pole. Given the fact that the pole is in the near field of the antenna it might affect its pattern and so the propagation of the signal.

\begin{figure}[!t]
    \centering
    \includegraphics[width=0.8\linewidth]{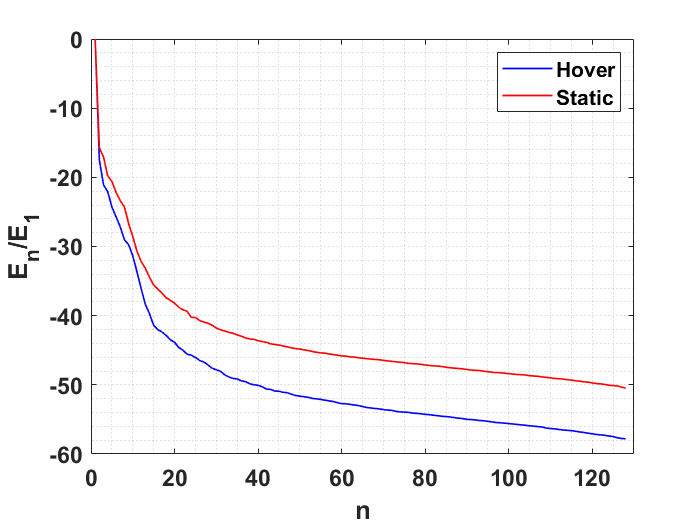}
    \caption{Sample of $R_i$ eigenvalue distribution for static and hover measurements.}
    \label{Ri}
\end{figure}

The RMS delay spread mainly depends on the type (LOS/NLOS) of the channel and the distance between TX and RX. Therefore, we expect similar results for both static and hovering given the fact that both are in LOS situations and located in the same place. The values for RMS delay spread are shown in Table \ref{tab:Ri_RMSDS} where $\bar{\sigma_\tau}$ is the mean value and $std(\sigma_\tau)$ is the standard deviation. The results are given on a dBs scale (as is common, e.g., in standardized channel models); the spread on a linear scale is  $10^{x/10}$, i.e. -90 dBs corresponds to 1 ns.
In case of the correlation matrix, our anticipation is that $\gamma_{12},\gamma_{14}$ will be large due to the fact that the measurement is in a LOS scenario and as consequence $R_i$ should be ill-conditioned. The results for the parameters $\gamma_{12},\gamma_{14}$ in dB scale are shown in Table \ref{tab:Ri_RMSDS}.
As can be seen in Table \ref{tab:Ri_RMSDS}, the values for $\gamma_{12},\gamma_{14}$ are at least $15$ dB, congruent to the LOS scenario analyzed. A sample eigenvalue distribution of the matrix $R_i$ is shown in Fig. \ref{Ri}. As expected, the range of the eigenvalues is approximately 50dB, indicating an ill-conditioned matrix.

\vspace{0.1in}
\begin{table}[]
\centering
\caption{Mean values for $\gamma_{12},\gamma_{14},\sigma_\tau$.}
\label{tab:Ri_RMSDS}
\begin{tabular}{|l|l|l|l|l|}
\hline
 & $\mathbf{\mu_{static}}$ & $\mathbf{\sigma_{static}}$ & $\mathbf{\mu_{hover}}$ & $\mathbf{\sigma_{hover}}$ \\ \hline
$\mathbf{\gamma_{12}[dB]}$ & 15.52 & 0.21 & 17.9  & 0.75 \\ \hline
$\mathbf{\gamma_{14}[dB]}$ & 19.76 & 0.14 & 22.32 & 0.83 \\ \hline
$\mathbf{\sigma_\tau[dBs]}$ & -78.52 & 0.08 & -79.44 & 0.27 \\ \hline
\end{tabular}
\end{table}

\subsubsection{Route}
\begin{figure}[!t]
    \begin{center}
    \subfloat[Power over the trajectory. ]{\includegraphics[width=0.8\linewidth]{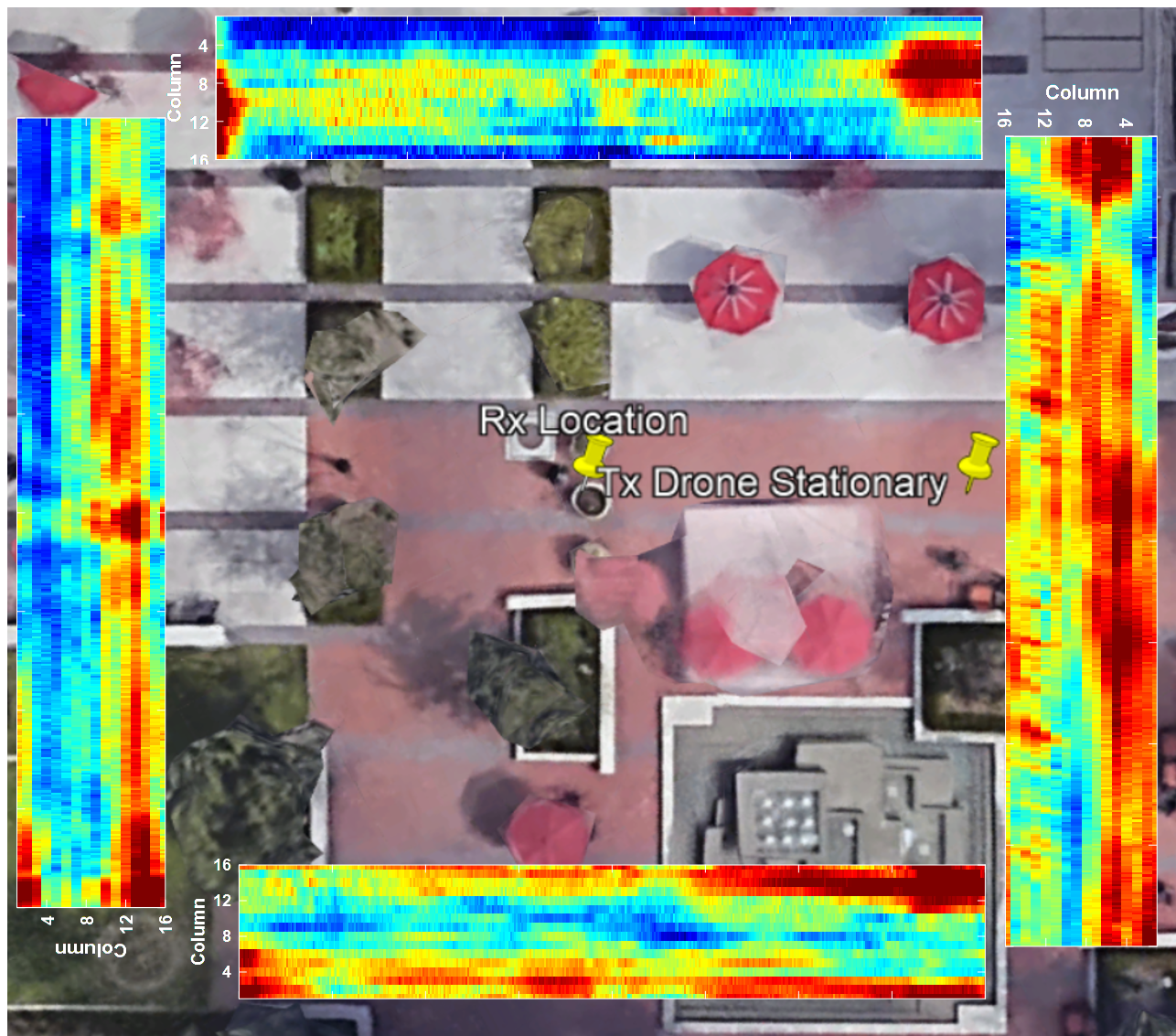}%
    \label{power_trajectory}}        
    \end{center}
    \begin{center}
    \subfloat[Full trajectory.]{\includegraphics[width=0.95\linewidth]{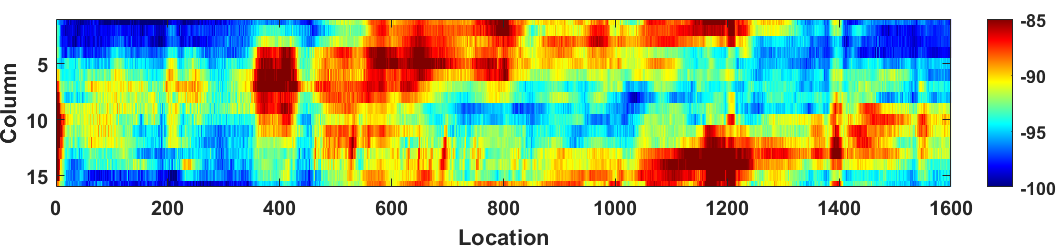}%
    \label{all_route_power}}        
    \end{center}

    \caption{Average power per column over the trajectory.}
    \label{route_power}
\end{figure}

Our analysis consists of checking the average power received by each column of the Massive MIMO array. Given the geometry of the scenario it can be anticipated that the antenna with the highest power will change between columns as we move the drone along the route as shown in \ref{power_trajectory}.\\
Figure \ref{all_route_power} shows the power per column given the locations. At the start of the route, columns 12 to 8 show the highest power given the fact that they are directly facing the receiver and also see reflections from the glass windows of the Olin Hall of Engineering (OHE) building. As the drone moves from east to west, the illumination of the building, and therefore the reflections from it, reduce and the LOS is the only strong component. When the drone goes south from the east edge of the route, we can see a large number of components coming from the windows of the OHE building. These MPCs reach the antenna columns facing back of the LOS direction. The third part of the route shows high power levels from columns 1 to 3 and 13 to 16 because they are facing the drone; additional MPCs come from reflections from the OHE building to columns 9 to 12 while the drone gets closer to the end of this part of the route. The last part of the route shows the columns 16 to 9 with the highest power as the drone moves to complete the route. In this case, the strongest MPCs come from the LOS directions and reflections on the windows of the OHE building.

\section{Conclusion}
In this paper, we presented the design and implementation of an air-to-ground directional massive MIMO channel sounder for drone-to-ground channels. Some of the features of our sounder are good phase and magnitude stability, and fast measurement rate, which are important for post-processing analysis. 
Test measurements were done to compare static and hovering drones, which allows to estimate the additional variance in the measurement from the drone. Our results have shown variations in the receiver power produced by the drone when it is hovering, effect deeply discussed in \cite{UAV_wob}. Additionally, given the number of antennas of our receiver and its cylindrical design a full coverage measurement can be obtained with a fine angular resolution, helpful for Fourier Analysis.
In our future work, extensive measurements will be conducted with the sounder in different environments.



\bibliography{references.bib} 
\bibliographystyle{IEEEtran}
\end{document}